# Galaxy Groups in Cold + Hot and CDM Universes: Comparison with CfA

**Richard Nolthenius**

UCO/Lick Observatory

University of California, Santa Cruz, CA 95064

**Anatoly Klypin**[1]

Dept. of Astronomy, New Mexico State University

Las Cruces, NM 88001

**Joel R. Primack**

Santa Cruz Institute for Particle Physics

University of California, Santa Cruz, CA 95064

## Abstract

This letter presents results of new high resolution $\Omega = 1$ Cold + Hot Dark Matter (CHDM) and Cold Dark Matter (CDM) simulations. Properties of groups in these simulations reflect the lower small-scale velocities and the greater tendency to form distinct filaments on both small and large scales in CHDM as compared to CDM. The fraction of galaxies in groups and the median group rms velocity are found to be powerful discriminators between models. We combine these two features into a very robust statistic, median group rms velocity $v_{\mathrm{gr}}(f_{\mathrm{gr}})$ as a function of the fraction $f_{\mathrm{gr}}$ of galaxies in groups. Using this statistic, we compare "observed" simulations to CfA data in redshift space in a careful and consistent way. We find that CHDM remains a promising model, with for example $v_{\mathrm{gr}}(0.45) \approx 125 \pm 25$ km s$^{-1}$ in agreement with the CfA data, while CDM with bias b=1.0 (COBE-compatible) or b=1.5, both giving $v_{\mathrm{gr}}(0.45) \approx 400 \pm 25$ km s$^{-1}$, can be virtually ruled out. Using median $M/L$, the observed value of $\Omega$ is 0.10 (CHDM) to 0.38 (CDM).

[1] On leave of absence from Astro-Space Center, Lebedev Physical Institute, Profsojuznaja 84, Moscow 117810, Russia

## 1. Introduction

A simple inflationary $\Omega = 1$ universe with a primordial Zeldovich spectrum of Gaussian, adiabatic fluctuations and a mixture of Cold + Hot Dark Matter (CHDM) has proven to be a serious candidate for the new, standard cosmological model (Klypin *et al.* 1993, hereafter KHPR), now that standard CDM seems unable to account simultaneously for small-scale galaxy pairwise velocities, large-scale motions and structure, and observed CMB anisotropies. Following KHPR, we consider a model with $\Omega_{\rm cold} = 0.6, \Omega_\nu = 0.3, \Omega_{\rm baryons} = 0.1$. The hot component is assumed to be a light neutrino, and specifying the model's single additional free parameter, the hot dark matter density fraction $\Omega_\nu$, fixes the neutrino mass. Interestingly, in the MSW scheme, the simplest and most attractive solution of the solar neutrino problem, the muon neutrino mass must be about 3 milli-eV; the 7 eV mass implied by $\Omega_\nu = 0.3$ is then in the range predicted for the tau neutrino by the simple (albeit speculative) "seesaw" model relating the masses of the neutrinos to those of the quarks (see e.g. Ellis *et al.* 1992).

When the CHDM initial fluctuation spectrum is normalized to the COBE $Q_{\rm ps-norm} = 17\mu{\rm K}$ (Smoot *et al.* 1992), corresponding to linear biasing parameter $b = 1.5$ ($\sigma_8 = b^{-1} = 0.67$), the CHDM model is found to compare favorably to observations on a number of measures, much better overall than CDM. These include small scale galaxy velocity dispersion and large scale streaming velocities (KHPR), cluster–cluster correlation function (Holtzman and Primack 1993, Klypin and Rhee 1993), power spectrum from CfA2 (Vogeley *et al.* 1992) and IRAS (Fisher *et al.* 1993, Feldman *et al.* 1993, Klypin *et al.* 1993b).

KHPR describe results of the analysis of particle-mesh (PM) simulations on a $256^3$ mesh for 14, 50, and 200 Mpc boxes (Hubble parameter $H_0 = 50$ km s$^{-1}$ Mpc$^{-1}$ is assumed throughout this paper). Klypin *et al.* (1993b) describes a more complete analysis of new, higher resolution CHDM and CDM simulations. Here, we focus on the most striking results from our study of properties of galaxy groups in these high resolution simulations, and compare them to identically selected groups in CDM simulations and in those formed from the CfA1 Survey (Davis *et al.* 1982).

## 2. Simulations, Galaxy Identification

The CHDM simulations were evolved from the initial power spectrum appropriate for the CHDM model at $z = 15$ to the present. Calculations were done with a PM code run on a $512^3$ mesh in a 100 Mpc box (i.e. cell size = 195 kpc) with $256^3$ cold particles ($M_{\rm particle} = 2.9 \times 10^9\ M_\odot$), and $2 \times 256^3$ hot particles ($M_{\rm particle} = 6.2 \times 10^8\ M_\odot$). For comparison, we also ran simulations of CDM (same mesh, $M_{\rm particle} = 4.1 \times 10^9 M_\odot$) at the same bias factor $b = 1.5$ (hereafter CDM1.5), and at the COBE-compatible CDM bias of 1.0 (hereafter CDM1). CDM1.5 was begun at $z = 18$, CDM1 at $z = 27.5$. Due to a statistical fluke of probability about 1/10, our first 100 Mpc CHDM simulation (hereafter, $CHDM_1$) had more power on the 100 Mpc wavelength than would be typical for a 100 Mpc box. The CDM simulations were run from the same choice of input waves. We also report here results from another simulation, $CHDM_2$, which had a more typical power spectrum for this box size. After running these simulations, we found that there were two mistakes in our initial conditions (see KHPR, Note Added in Proof): the fitting formula for the cold spectrum was too small and the velocities were too large, both by about 20% on small scales. However, these effects are in phase and largely cancel. We have run another $CHDM_2$ (rev) simulation with both errors corrected, and found that the power and velocity differences between this and $CHDM_2$ (old) on small scales declined to 5% by $z = 7$ and remained at this level. All results discussed here are from $CHDM_2$ (rev), but the agreement with $CHDM_2$ (old) is well within the $1\sigma$ error bars; thus the $CHDM_1$ and CDM simulations should also be reliable.

Dark halos ("galaxies") in our simulations were identified as local density maxima on the $512^3$ mesh, and assigned the mass inside this cell. Only dark halos with mass $M > 4.1 \times 10^{10} M_\odot$, corresponding to $\delta\rho/\rho > 80$, were used in the analysis below.

### 3. Sky Catalogs, Group Identification

Group properties reflect features of the CHDM and CDM models on the $\sim 0.1 - 10$ Mpc scales where they differ most significantly. In order to make reliable comparisons with observed groups, it is essential to "observe" simulations and identify groups in redshift space in exactly the same way as is done for the data. Since our simulation resolution of 195 kpc per cell is more coarse than observations, we began by merging CfA1 grouped or paired galaxies from Nolthenius (1993; N93) which were closer than a sky projected separation of 235 kpc. This merged catalog had 2112 galaxies, compared to the original 2406. 195 kpc is only $\sim 10\%$ of the median group harmonic radius $\langle r_h \rangle_{\mathrm{med}}$, and our limited resolution should not seriously affect our results. (See NW for definitions of $\langle r_h \rangle_{\mathrm{med}}$ and other quantities used here.)

We assigned luminosities $L$ to halo masses in two different ways. The first and in our opinion more reliable prescription was to randomly generate Schechter-distributed magnitudes, using merged CfA1 parameters $\alpha = -1.17$ and $M^* = -21.25$, and assign these to halos with luminosity monotonically increasing with halo mass (Schechter method). We also tried using the relation between dark matter density and baryonic density found in hydrodynamic simulations by Cen and Ostriker (1992, 1993), assuming constant $M_{\mathrm{baryon}}/L$ (CO method). (We used the masses in the $3^3$ cells centered on each halo for $\hat{\rho} \equiv \rho/\bar{\rho}$ in CO's relation $\log(\hat{\rho}_{\mathrm{baryon}}) = A + 2.3 \log(\hat{\rho}_{\mathrm{tot}}) - 0.20 \log^2(\hat{\rho}_{\mathrm{tot}})$, and took $A = 5.8$ for CDM and $A = 6.23$ for CHDM simulations to match the merged CfA1 galaxy density.) The CO method produces a luminosity function which is too steep at low luminosities ($\alpha \simeq -1.8$, essentially the Press-Schechter result) and has too many bright galaxies. This weighted the sample in favor of distant, bright (and to a lesser extent, nearby and faint) galaxies much more heavily than the CfA data, and moderately raised median group sizes, rms velocities, and fractions grouped. While we regard the CO $L(M)$ as unrealistic, it turned out to give essentially the same results for our favored statistic $v_{\rm gr}(f_{\rm gr})$ as our preferred Schechter method.

We partially compensate for cosmic variance by selecting six viewing locations which approximate our place in the local universe, using the same viewing location criteria for both CHDM and CDM simulations: (a) after applying a magnitude limit (using the Schechter method), halo density within 15 Mpc in redshift space was within a factor 1.5 of CfA1 density, and (b) a roughly Virgo-mass small cluster was $\sim 25$ Mpc distant. (By using the same random numbers for the $CHDM_1$ and CDM simulations, we insured that the relative halo densities and locations of large concentrations were virtually the same, and we could then place CDM viewers on the nearest halo to the $CHDM_1$ viewers.) Assigned magnitudes were then cut at $m = 14.5$, and the resulting catalogs randomly culled an additional $\sim 20\%$ down to CfA1 density. Finally, in order to include edge effects, a 20 degree wide "zone of avoidance" was added. This produced sky catalogs of 10.38 sr, compared to the 2.66 sr of the CfA1 sample.

We then used the adaptive friends-of-friends link algorithm of N93 to identify groups. Each galaxy is surrounded by a cylinder with its axis along the line of sight, and three or more touching or intersecting cylinders are considered to define "groups." The radius of the cylinder is a constant fraction $D_n$ ("sky link" parameter) of the mean galaxy spacing on the sky, and the length is $V_L(d)$ ("redshift link," parameterized by its value $V_5 \equiv V_L(5000)$ at the typical CfA1 distance $d = 5000$ km s$^{-1}$). Both sky link and redshift link increase with sample incompleteness (and therefore distance) according to the fitted luminosity functions for each catalog as discussed in N93. A fiducial $D_n = 0.36$ (1.2 Mpc at distance 1000 km s$^{-1}$) produces groups of minimum overdensity about 20 and is a good compromise between selecting only dense cores or percolating connections between groups on the sky; it is quite similar to that used in earlier studies (Geller and Huchra 1983; Nolthenius and White 1987 (NW); Ramella, Geller, & Huchra 1989; N93; and Moore *et al.* 1993 (MFW); note that NW, MFW, and N93 express $D_n$ as a fraction of the 3-D particle spacing, rather than the more appropriate projected spacing used here, so MFW's fiducial $D_0 = 0.3$ corresponds to our $D_n = 0.3 \times \pi/2 = 0.47$).

## 4. Results

The primary results of this study follow from two findings: CHDM simulations are both much more filamentary and also have lower small-scale velocities than the CDM simulations. This means that groups form and percolate much more readily in CHDM, both in real and redshift space. Nevertheless, we find that the full distributions of multiplicity, mass, velocity dispersion, and size near the fiducial links are poor discriminators between models; all our simulations are similar, and, within the errors, in reasonable agreement with the CfA1 data.

NW first demonstrated the value of comparing the median/percolation properties of groups versus link parameters, finding that CfA1 groups form and percolate more readily than in low density b=1.0 CDM. Here we extend this idea and present a sensitive and robust statistic for discriminating between model simulations. Note that for most purposes our conclusions do not require that "groups" actually be dynamically isolated objects, only that they be consistently identified. We may therefore make valid comparisons througout the link range, even though conventionally defined groups are only approximated near the fiducial links. In this sense "groups" are only a label we apply to a set of objects which provide useful discriminating statistics.

Figure 1 compares the fraction of galaxies $f_{\rm gr}$ found in groups vs. redshift link $V_5$ at our fiducial sky link $D_n = 0.36$. $f_{\rm gr}$ rises steeply at small $V_5$ as valid group members are added to groups. Within this regime, median group rms velocities $v_{\rm gr}$ are artificially clipped to that allowed by $V_5$; we refer to this as the "clipped regime". At higher $V_5$, only interlopers from the lower density surroundings are picked up, and the curves flatten ("interloper regime"). This transition is rather distinct for the CHDM and CfA data. This is consistent with our visual inspection of the halo distribution: we found that CHDM halos are mostly found in filaments and knots, much more so than in the CDM boxes, which show a more gradual transition to the interloper regime. There is remarkably little variation between the six different viewpoints, as shown on the simulation points on all the figures by the $1\sigma$ error bars. For the CfA points, the errors were estimated from the $CHDM_1$ simulation; we took the rms spread among its six viewpoints after restricting each viewpoint's sky to the same 2.66 sr divided into two regions as for CfA1, one region including the nearby Virgo-sized cluster. Such error estimates include both statistical scatter and cosmic variance, but the difference between the two CHDM simulations shows that these error bars do not include all the cosmic variance. It is clear that both CDM1 and CDM1.5 seriously conflict with the data.

Next we consider how the median group rms velocity $v_{\rm gr}$ depends on the redshift link. (We define the rms velocity of each group as $v_{\rm gr} = [(n-1)^{-1} \sum_i (v_i - <v>)^2]^{1/2}$ = "group velocity dispersion" in Geller and Huchra 1983; here $v_i = cz_i$.) Figure 2 shows $v_{\mathrm{gr}}$ as a function of $V_5$, here done at the larger $D_n = 0.47$ to facilitate comparison with MFW. Similar to Figure 1, all curves show a steeper rise in the clipped regime followed by a shaller rise in the lower density interloper regime. A similar curve at our fiducial $D_n = 0.36$ shows CHDM transitions at $V_5 \simeq 350$ km s$^{-1}$, and CDM at $V_5 \sim 600$ km s$^{-1}$. The curves show that within the clipped regime group rms velocities are limited by the $V_5$ link rather than the underlying properties of the simulation. The transition to a flatter slope is slower in the CDM curves of Figures 1 and 2, indicating a more substantial tail of high velocity neighbors. CDM $v_{\mathrm{gr}}$'s are significantly and consistently higher than the data even at low $V_5$.

For our conclusions, it is important to be sure we have not overestimated CDM small scale velocities. MFW used the CDM simulations of Frenk *et al.* (1990) to study galaxy groups. Their P$^3$M code gave better spatial resolution than our PM code, and their galaxy identification scheme ("high peaks"), galaxy luminosity function, and link algorithm (NW) also differ from ours. Nevertheless, our $v_{\mathrm{gr}}$ for CDM1.5 at the same $D_n$ compares well with their CDM b=1.6 results, as shown in Figure 2. (No error bars are shown on the MFW points, since MFW did not estimate them.)

Figures 1 and 2 suggest that the optimal $V_5$ for producing the largest relatively uncontaminated groups is near $V_5 \simeq 350$ km s$^{-1}$ for CHDM groups and 600 km s$^{-1}$ for CDM groups. To confirm this, we made groups from these same sky catalogs using full 3D information with a 3D link parameter $D_n = 0.36$. Reproducing the 3D group's $v_{\mathrm{gr}}$'s required a $V_5 \simeq 350$ km s$^{-1}$ for both CHDM simulations, and a $V_5 = 450$ km s$^{-1}$ (CDM1.5) to 550 km s$^{-1}$ (CDM1) for CDM. Matching fractions grouped required slightly higher $V_5$'s. We cover this range by adopting fiducial $V_5$ of 350 for CHDM and 600 km s$^{-1}$ for CDM. Note that $V_5 = 350$ km s$^{-1}$ also gives the optimum CfA1 groups (NW, N93).

A statistic that is both discriminatory and robust is obtained by plotting the median group rms velocity as a function of the fraction of galaxies in groups. Since raising $V_5$ typically raises both $f_{\mathrm{gr}}$ and $v_{\mathrm{gr}}$, plotting $v_{\mathrm{gr}}(f_{\mathrm{gr}})|_{D_n}$ at fixed $D_n$ (i.e. eliminating $V_5$) produces a strong discriminator between models, shown in Figure 3. From left to right, Figure 3 points correspond to $V_5$ from 40 to 2000 km s$^{-1}$. Note that through its implicit dependence on $V_5$, this plot is sensitive to differences in the small-scale velocity dispersion between models. This statistic turns out to be quite robust. Using the very different Cen-Ostriker luminosity assignment produces virtually identical curves, as does a change from $D_n = 0.36$ to 0.47 (though here all curves are pushed to slightly higher $f_{\mathrm{gr}}$). It does not appear possible to reconcile CDM groups with the CfA data for any reasonable selection parameters or mass-to-light assignment. Over most of the range, CHDM agrees well with the CfA data. Note also that, to the extent the CfA region is typical, the CHDM$_2$ initial conditions are a more appropriate comparison. On all figures, CHDM$_1$ (and associated

CDM curves as a block) should then be moved towards the CHDM$_2$ curves. This worsens CDM's and improves CHDM's agreement with the data.

We also considered a related statistic $v_{\rm gr}(f_{\rm gr})|_{V_5}$, this time varying the sky link parameter $D_n$, keeping $V_5 = 600$ km s$^{-1}$ fixed (see Klypin *et al.* 1993b). Again, CDM $v_{\rm gr}$'s are $50 - 100$ km s$^{-1}$ too high and clearly inconsistent with the data. CHDM curves agree much better, but are moderately too high at small $f_{\rm gr}$ (corresponding to small $D_n$).

Finally, we determined the observed value of $\Omega$ from the simulations. Using the usual $M/L$ method (e.g. NW), we found $\Omega$'s of 0.10 (both CHDM), 0.17 (merged CfA1), 0.30 (CDM1.5) and 0.38 (CDM1). CDM $\Omega$'s are higher largely due to their higher fiducial $V_5 = 600$ km s$^{-1}$. These demonstrate that it is indeed possible to measure low $\Omega$ from groups in an $\Omega = 1$ universe.

## 5. Caveats and Conclusions

While CHDM appears to do much better than CDM in these comparisons to observations, there remain discrepancies that appear significant. In particular, at small $D_n$ we are looking at dense group cores, which appear to be significantly cooler in the CfA data than any of the simulations. At the smallest $D_n = 0.15$, CHDM $v_{\rm gr}$'s are almost 50 km s$^{-1}$ higher than the data. Also, CfA $v_{\rm gr}$'s show a significantly steeper rise with $f_{\rm gr}$ than either CHDM or CDM. A related point is that simulation groups show larger median group harmonic radius $\langle r_h \rangle_{\rm med}$ than those observed, by $\sim 20\%$ for CHDM, $\sim 30\%$ for CDM, especially at small $V_5$ and $D_n$. These facts suggest that real group cores are both colder and denser than in CHDM and especially CDM simulations. The additional 10% discrepancy seen in CDM is likely to be genuine, a result of higher pairwise velocities inflating the groups. For CHDM, the discrepancy may be due in part to our limited resolution at these scales ($D_n = 0.15 \simeq 0.75$Mpc $\simeq 3 - 4$ cells for nearby groups) or the fact that dissipation is not modelled, or may indicate a real difference. At least some of the $\langle r_h \rangle_{\rm med}$ discrepancy is likely due to the problem of the "overmerging" of dark matter halos in dissipationless simulations (e.g. Gelb and Bertschinger 1993), which is undoubtedly present in our simulations. Overmerging will tend to remove close neighbors and thus inflate average pairwise spacings. In CDM models this appears significant for halo masses $M > 10^{12} M_\odot$ (Evrard, Summers, & Davis 1993), which comprise 40% of the halos in our CDM models, most of these in groups. However, overmerging actually lowers pairwise velocities, so any correction would most likely worsen CDM's already desperate plight with respect to group rms velocities. For CHDM, the hot component delays collapse epochs and thus helps reduce overmerging, and while no studies comparable to Evrard, Summers, & Davis suggest at what mass it becomes significant, only $21 - 28\%$ of our CHDM halos are above $10^{12} M_\odot$. Thus, correcting for overmerging is likely to widen the differences between CDM and CHDM, and between CDM and the data, thereby strengthening our basic conclusions.

To summarize our conclusions: Median group rms velocity $v_{\rm gr}$ versus fraction of galaxies in groups $f_{\rm gr}$ is shown to be a powerful and robust statistic for discriminating between cosmological models. Both unbiased and low bias $\Omega = 1$ CDM groups show unacceptably high median rms velocities and too few galaxies in groups, due to the combined effects of high pairwise velocities and less tendency for halos to be arrayed in filaments and knots. CHDM's highly filamentary nature percolates more easily and makes groups slightly easier than observed in the CfA1 data. CHDM remains a promising model, with $v_{\rm gr}(0.45) \approx 125 \pm 25$ km s$^{-1}$ at $D_n = 0.36$ in agreement with the CfA1 data. At our fiducial $D_n$ and $V_5$, $f_{\rm gr} = 0.42$ for merged CfA1 and there are 170 groups. A bootstrap estimate of the purely statistical error in $v_{\rm gr}(0.45)$ for CfA1 is 9%; the rms among the six viewpoints is 11%, giving $v_{\rm gr}(0.45) = 138 \pm 15$ km s$^{-1}$. CDM with bias b=1.0 (COBE-compatible) and b=1.5, both with $v_{\rm gr}(0.45) \approx 400 \pm 25$ km s$^{-1}$ at $D_n = 0.36$, can be virtually ruled out.

The $v_{\rm gr}(f_{\rm gr})$ statistic should be measured for the larger redshift data sets that will soon become available (e.g. CfA2, SSRS2, Perseus-Pisces), and also on high-resolution simulations of all the potentially viable cosmological models, including CHDM with varying fractions of $\Omega_\nu \approx 0.2 - 0.25$, and $\Omega_{\rm baryon} \approx 0.05$. It will be interesting to see which models can pass this difficult test!

**Acknowledgments**. We thank J. Ostriker for suggesting that we use CO for assigning luminosities, and the Aspen Center for Physics for its hospitality in August 1993. Our simulations were done on the Convex-3880 supercomputer at the NSCA.

# References

Cen, R. & Ostriker, J.P. 1992, Ap J, 399, L113

Cen, R. & Ostriker, J.P. 1993, Princeton preprint (CO)

Davis, M., Huchra, J., Latham, D.W., and Tonry, J. 1982, Ap J, 253, 423

Ellis, J., Lopez, J. L., & Nanopoulos, D. V. 1992, Phys. Lett. B, 292, 189

Evrard, A., Summers, F., and Davis, M. 1993, preprint

Feldman, H., Kaiser, N., & Peacock, J. A. 1993, preprint

Fisher, K.B. *et al.* 1993, Ap J, 402, 42

Gelb, J., and Bertschinger, E., 1993, Ap J, submitted

Geller, M.J., & Huchra, J.P., 1983, Ap J Supp, 52, 61

Holtzman, J.A. & Primack, J.R. 1993, Ap J, 405, 428

Klypin, A., Holtzman, J.A., Primack, J., & Regos, E. 1993a, Ap J, 416, 1 (KHPR)

Klypin, A., Nolthenius, R., & Primack, J., 1993b, in preparation

Klypin, A., & Rhee, G. 1993, Ap J, in press

Nolthenius, R.A. 1993, Ap J Supp, 85, 1 (N93)

Nolthenius, R.A., & White, S.D.M. 1987, M.N.R.A.S., 225, 505 (NW)

Moore, B., Frenk, C.S., and White, S.D.M. 1993, M.N.R.A.S., 261, 827

Smoot, G.F. *et al.* 1992, Ap J, 396, L1

Vogeley, M.S., Park, C., Geller, M.J. & Huchra, J.P. 1992, Ap J Lett, 391, 5

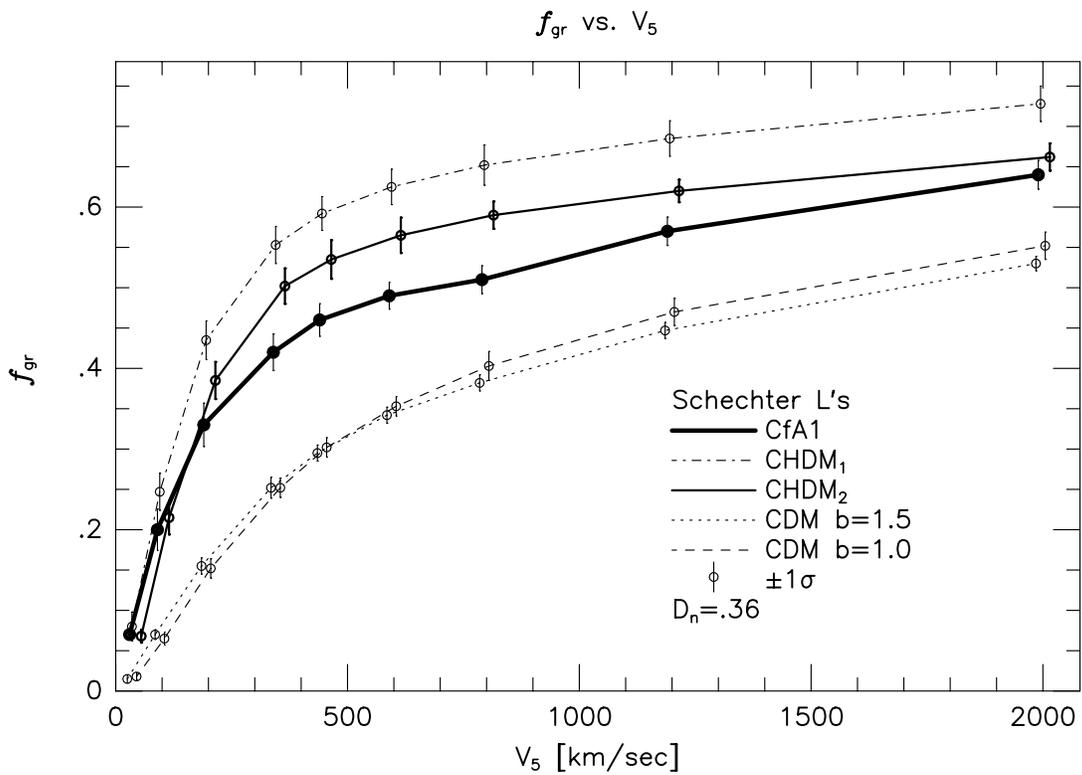

Figure 1. The fraction of galaxies grouped rises sharply at low redshift link $V_5$ as valid group members are added, then levels off as only interlopers are included. CDM consistently groups a lower fraction than the data, and shows a less distinct transition to the interloper regime at higher $V_5$. CHDM groups slightly too high a fraction.

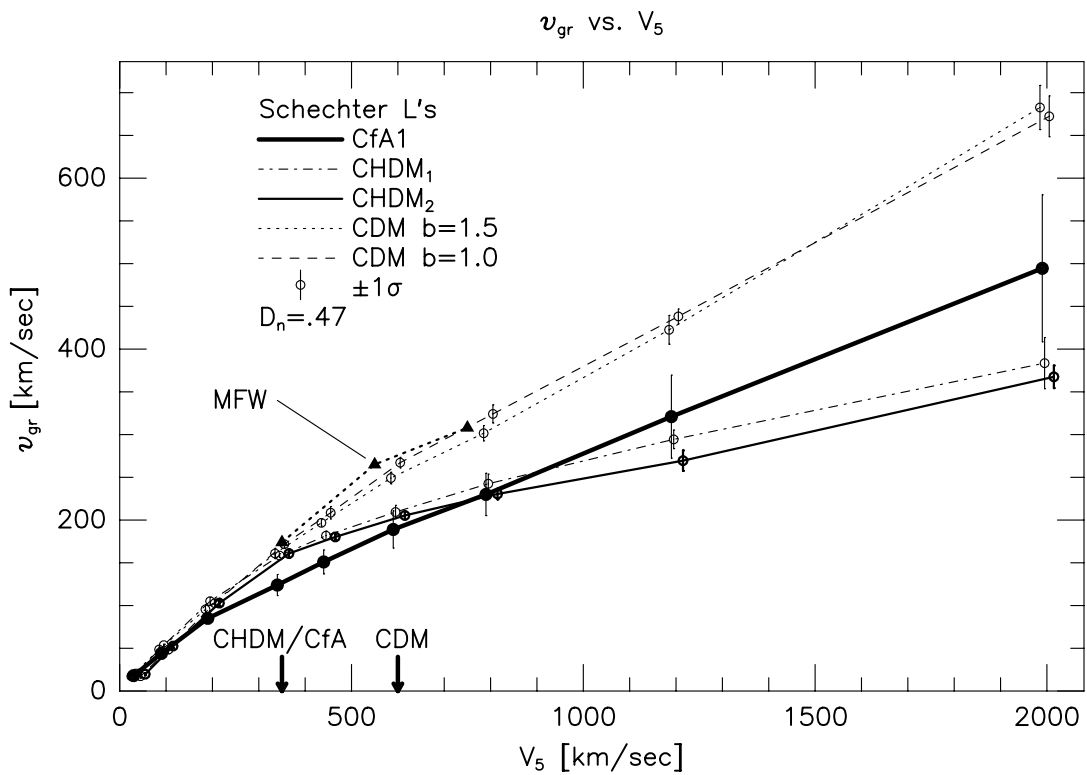

Figure 2. $v_{\rm gr}$ is here plotted vs. $V_5$ at MFW's fiducial $D_n$, with MFW's CDM results also plotted. $v_{\rm gr}$ rises more steeply at low $V_5$, then flattens somewhat for the same reasons as in Figure 1. At low $V_5$, rms velocities are artificially clipped for all simulations. Our fiducial $V_5$'s (at our fiducial $D_n = 0.36$) are marked with arrows. Beyond the clipped regime, CDM rms velocities are consistently too high.

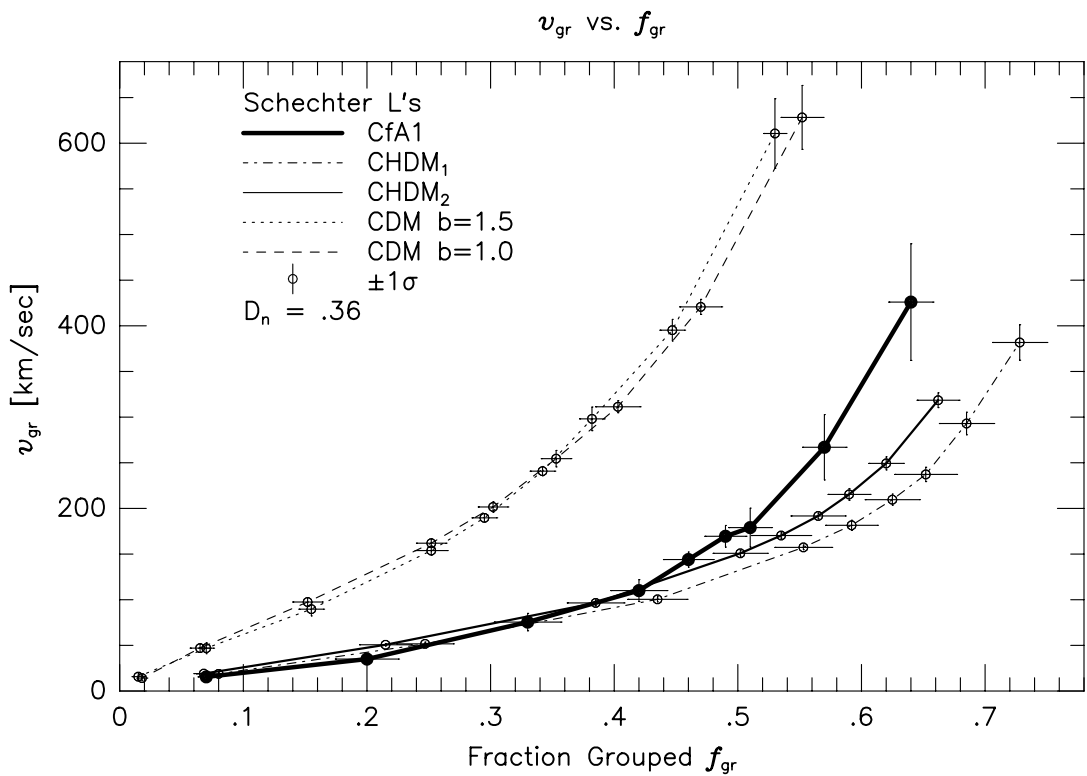

Figure 3. Median group rms velocity $v_{\rm gr}(f_{\rm gr})|_{D_n}$ of galaxies in groups at fixed sky link $D_n$ proves to be a powerful and robust statistic. It is plotted here for our fiducial sky link $D_n = 0.36$, and Schechter luminosity assignments. CDM simulations for both bias choices produce $v_{\rm gr}$ far higher than CfA1 groups, while both CHDM simulations are in good agreement. Relative positions of curves remain virtually unchanged under changes in sky link $D_n$ or luminosity assignment method.